\documentclass[a4paper]{jpconf}
\usepackage{graphicx}

\newcommand{\pubjournal}[4]{#1, #2, #3 (#4)}

\begin{document}

\title{The H.E.S.S. View of the Central 200 Parsecs}

\author{Jim Hinton$^{1,2}$ for the H.E.S.S. Collaboration}

\address{$^1$ Max-Planck-Institut f\"ur Kernphysik, P.O. Box 103980, D 69029 Heidelberg, Germany}
\address{$^2$ Landessternwarte, Universit\"at Heidelberg, K\"onigstuhl, D 69117 Heidelberg, Germany}
\ead{Jim.Hinton@mpi-hd.mpg.de}

\begin{abstract}
The inner few hundred parsecs of our galaxy provide a laboratory for
the study of the production and propagation of energetic
particles. Very-high-energy $\gamma$-rays provide an effective probe
of these processes and, especially when combined with data from other
wave-bands, $\gamma$-rays observations are a powerful diagnostic
tool. Within this central region, data from the H.E.S.S. instrument
have revealed three discrete sources of very-high-energy $\gamma$-rays
and diffuse emission correlated with the distribution of molecular
material. Here I provide an overview of these recent results from
H.E.S.S.
\end{abstract}

\section{Introduction}

The central 200~pc of our Galaxy is a unique region that harbours many
remarkable objects --- including several potential sites of effective
particle acceleration. Non-thermal emission (particularly in the
radio, X-ray and $\gamma$-ray bands) can be used to trace the
energetic particle populations of this region. X-ray and radio
observations of synchrotron emission provide information on the
product of the local magnetic field energy density and the density of
relativistic electrons. The flux of inverse Compton $\gamma$-rays, on
the other hand, is proportional to the radiation field density (and the
electron density). The combination of $\gamma$-ray and X-ray
measurements therefore provides a powerful tool for probing both
magnetic field strength and energetic particle content. Moreover,
$\gamma$-rays provide an effective tracer for hadronic particles:
proton-proton interactions in the interstellar medium lead to the production
and decay of pions and hence $\gamma$-ray production. The combination
of $\gamma$-ray measurements with tracers of atomic and molecular
material may be the \emph{only} way to effectively trace energetic
hadrons in our galaxy.

The usefulness of $\gamma$-ray observations has traditionally been
limited by poor angular resolution and modest sensitivity.  A major
step forward in the very-high-energy (VHE, $>100$ GeV) domain has
recently been taken with the commissioning of H.E.S.S.  (High Energy
Stereoscopic System). H.E.S.S. is an array of four, 13~m diameter,
imaging Cherenkov telescopes located in the Khomas highlands of
Namibia~\cite{HESS}, a southern hemisphere location ideal for
observations of the Galactic Centre (GC).  H.E.S.S. has an angular
resolution of a few arc-minutes and a locational accuracy of $\sim
30''$ for typical point sources. The instrument reaches an energy flux
sensitivity of $10^{-12}$ erg cm$^{-2}$ s$^{-1}$, an order of magnitude
lower than the previous generation of VHE instruments.  The wide field
of view of H.E.S.S. (5$^{\circ}$ in diameter) enables us to
simultaneously monitor the entire central 200 parsec region. Results
from the first two years of H.E.S.S. GC observations are described
here.

\begin{figure}[h]
  \begin{center}
    \includegraphics[width=36pc]{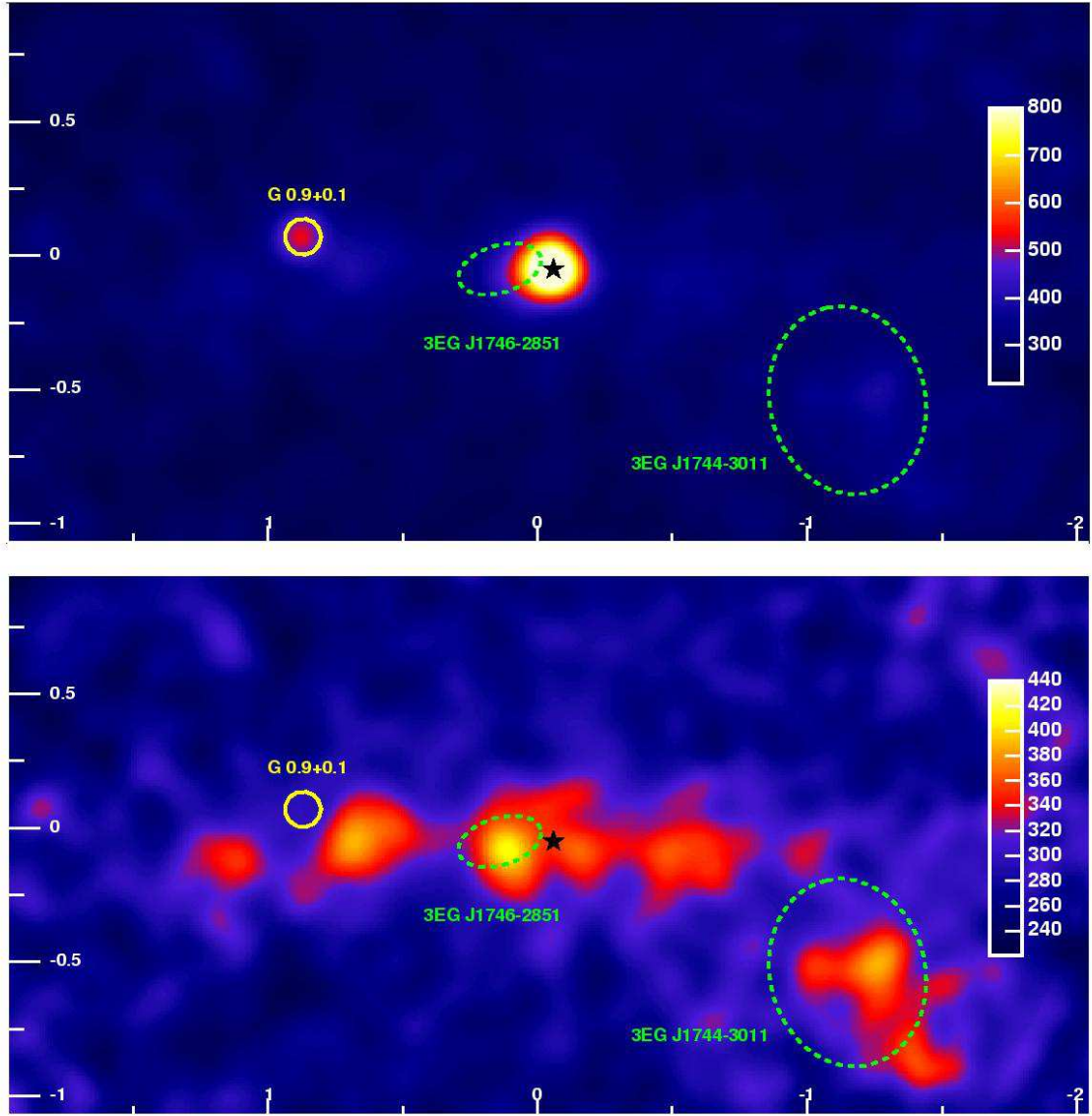}
    \caption{\label{fig1} The H.E.S.S. view of the central 200 parsecs
      (reproduced from \cite{HESS_gc}).  Top panel: smoothed count map
      (without background subtraction) for data taken with H.E.S.S. in
      2004.  Bottom panel: the same data after subtraction of
      point-like excesses at the positions of Sgr A$^{\star}$ and
      G0.9+0.1. 
      The location of Sgr~A$^{\star}$ is
      marked with a black star and the G\,0.9+0.1 as a yellow circle.
      95\% confidence regions for the positions of EGRET sources are
      shown as dashed ellipses~\cite{EGRETEllipses}.  }
  \end{center}
\hspace{2pc}
\end{figure}

\section{The central VHE gamma-ray source: HESS J1745-290}

During 2004, detection of a compact source of VHE $\gamma$-rays close
to the GC was claimed almost simultaneously by groups using three
Imaging Air-Cherenkov Telescope (IACT) systems: CANGAROO~\cite{CANGAROO},
Whipple~\cite{VERITAS} and the partially complete
H.E.S.S. array~\cite{HESS_saga}.  The best localisation of this source
(HESS J1745-290) is that derived from H.E.S.S. data, which places 
the emission within $1'$ of Sgr A$^{\star}$.
The initial spectral results from the three groups were originally in
serious disagreement. While all groups agree on a power-law spectrum
($F\propto E^{-\Gamma}$), the photon indices of H.E.S.S. ($\Gamma =
2.21 \pm 0.09$) and CANGAROO ($\Gamma =4.6 \pm 0.5$) differ
substantially, and the single flux point at 2.8~TeV published by the
Whipple collaboration was a factor 1.5 above the
H.E.S.S. flux. However, further analysis of the CANGAROO data yielded a
much larger error on the photon index: $4.6^{-1.2}_{+5.0}$
\cite{CANGAROO_VELAJNR} and a more detailed analysis of the Whipple
data yielded a photon index of $2.44^{+0.22}_{-0.48}$ and a flux
consistent with the H.E.S.S. value~\cite{KosackPhD}.  The final
confirmation of the hard energy spectrum seen by H.E.S.S.  came with
the detection of this source by MAGIC~\cite{MAGIC}.  The MAGIC flux
and photon index ($2.2 \pm 0.2$) show good agreement with
H.E.S.S. values.

The construction of the H.E.S.S. array was completed in late 2003 and
50 hours of full sensitivity observations of the GC took place in
2004. The upper panel of Fig.~\ref{fig1} shows a smoothed count map of
the central four degrees of our galaxy from these 2004
data~\cite{HESS_gc}.  Two VHE $\gamma$-ray sources are clearly
apparent: one coincident with the SNR G\,0.9+0.1 (discussed in
section~\ref{secG09}) and the unidentified source HESS J1745-290 in
the Sgr A region. The higher sensitivity observations in 2004 confirm
the spectral shape and source position derived from the
H.E.S.S. two-telescope 
data.  The source position derived from the 2004 dataset
lies $5''\pm10''_{stat}\pm20''_{sys}$ from Sgr A$^{\star}$.  The
updated photon index is $\Gamma = 2.29\pm0.05_{stat}\pm0.15_{sys}$
\cite{HESS_icrc}. No significant variability is found on time-scales of
days or hours in the 2004 dataset. 

Within the error circle of the H.E.S.S measurement lie three
compelling candidates for the origin of the VHE emission: the
shell-type supernova remnant (SNR) Sgr~A~East~(see e.g.~\cite{AEast},
the newly discovered pulsar wind nebula G\,359.95$-$0.04~\cite{wang}
and the supermassive black hole Sgr~A$^{\star}$ itself~(see
e.g.~\cite{AharonianSAGA}).  Several radiation mechanisms have been discussed
for these three objects. Plausible mechanisms include inverse Compton
scattering of energetic electrons, the decay of pions produced in the
interactions of energetic hadrons with the interstellar medium or dense radiation
fields and finally curvature radiation of ultra-high energy protons
close to Sgr A$^{\star}$.  For some production scenarios, correlated
variability is expected in X-rays and $\gamma$-rays.  Simultaneous
observations of the Sgr~A region with H.E.S.S. and Chandra took place
in July 2005 and may help to clarify this situation.

%

A widely discussed alternative to these astrophysical origins is
the annihilation of dark matter in the central cusp in the density profile of
our galaxy. However, the hard power-law spectrum of the central source
is hard to reconcile with a dark matter interpretation
(see~\cite{GCDM} for a detailed discussion).

\begin{figure}[h]
\begin{center}
\includegraphics[width=32pc]{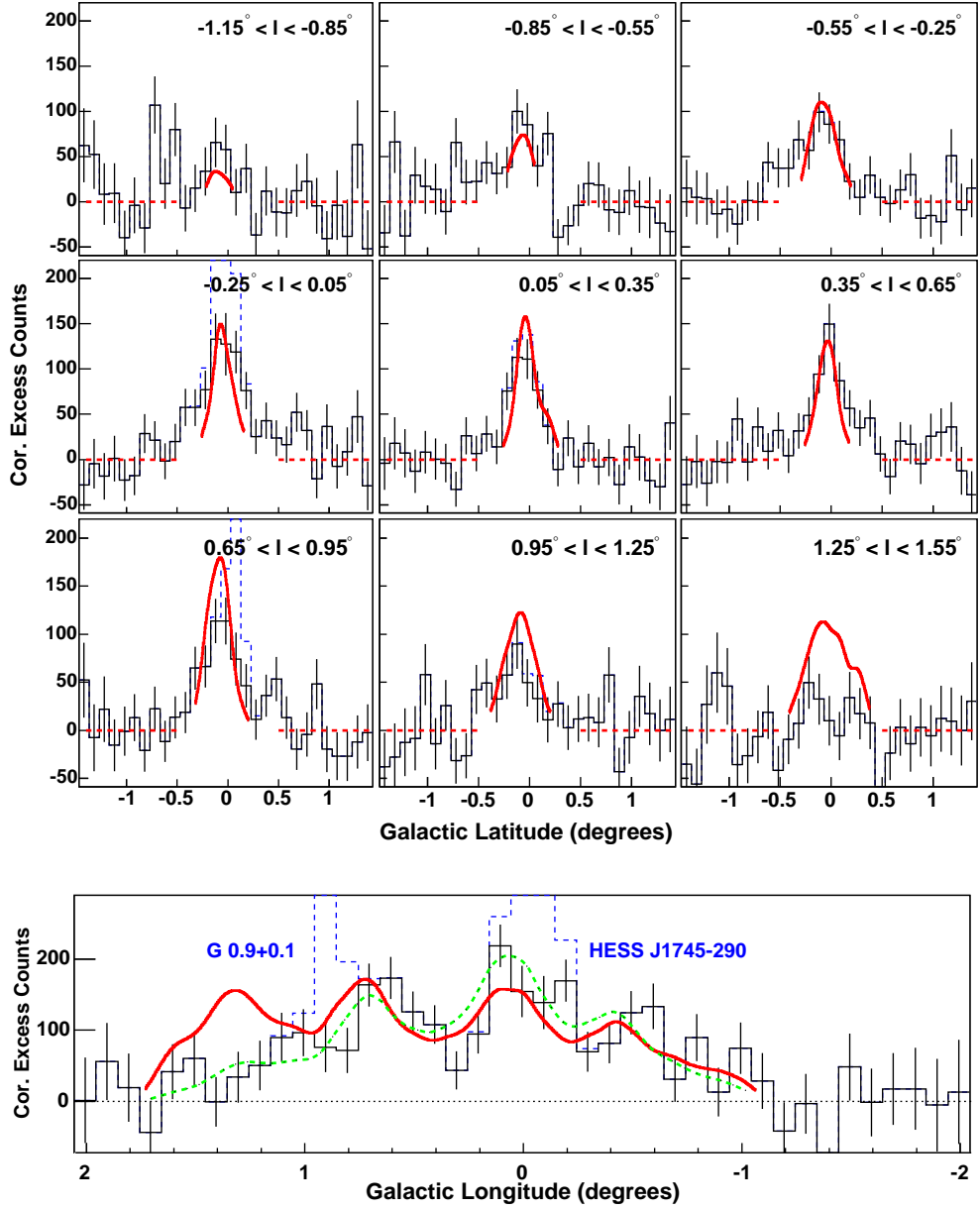}
\end{center}
\hspace{2pc}
\caption{\label{fig2} Slices through the VHE $\gamma$-ray emission of
  the galactic centre region. Top panels: latitude slices in different
  longitude bands.  Bottom panel: longitude slice for $\pm0.2^{\circ}$ in latitude.
  The dashed (solid) histograms show the $\gamma$-ray emission before
  (after) the subtraction of point-like sources at the positions of
  G\,0.9+0.1 and Sgr\,A$^{\star}$.  The red curves show slices through
  smoothed CS emission maps from \cite{Tsuboi}. The dashed green curve
  in the lower panel shows the expected $\gamma$-ray emission for a
  population of CRs injected $\sim10^{4}$ years ago, assuming nominal
  'disc-like' diffusion (figure reproduced from \cite{HESS_gc}) }
\end{figure}

\section{Diffuse emission from the Central Molecular Zone}

\begin{figure}[h]
\begin{center}
\includegraphics[width=30pc]{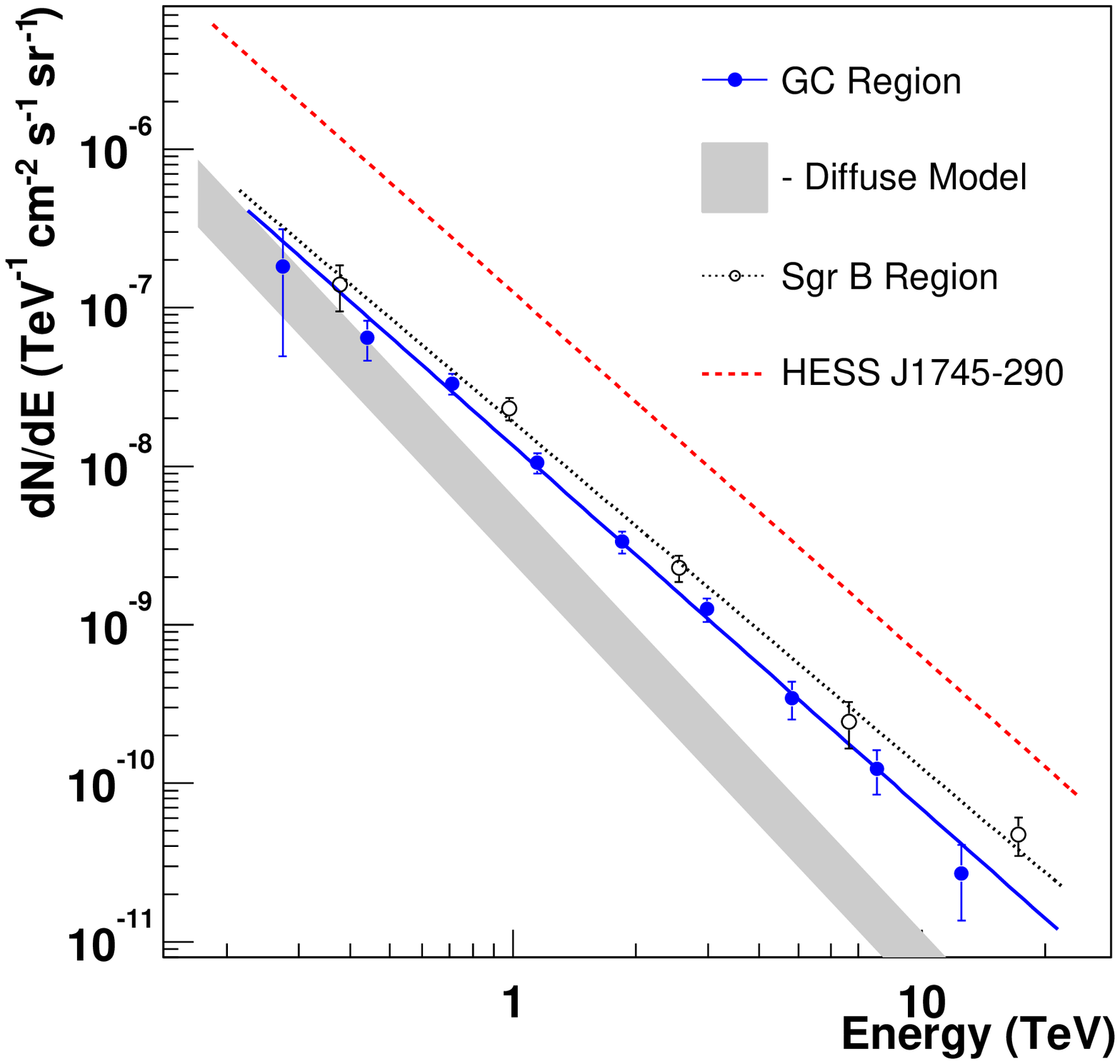}
\end{center}
  \caption{\label{fig3} The differential energy spectra of the diffuse
    GC emission (solid line), the Sgr B molecular complex (dotted
    line) and the central source HESS J1745-290 (dashed line).  The
    grey shaded box illustrates the expected flux for the GC region
    from the interactions of CRs with a spectrum identical to that
    measured at the Earth. The width of the box indicates the
    uncertainty on the total molecular mass of the CMZ (figure reproduced
    from \cite{HESS_gc}).  }
\end{figure}

The dense gas clouds of the central molecular zone provide
effective targets for the production of $\gamma$-rays via the
interactions of hadrons. Indeed, the recent H.E.S.S. observations
reveal TeV emission correlated with these clouds (see the lower panel
of Fig.~\ref{fig1}).  This is the first case in which such a
correlation has been seen at these energies.  Galactic longitude and
latitude distributions of the diffuse VHE emission are shown in
Fig.~\ref{fig2}.  In Fig.~\ref{fig2}, maps of
CS (J=1-0) emission from \cite{Tsuboi} have been used to estimate the
column density of the GC clouds. The ratio of $\gamma$-ray emission to
column density provides a measure of the density of TeV cosmic rays (CRs)
in the Galactic Centre region. Assuming that the locally measured CR
spectrum is valid in the GC, we can predict the expected
$\pi^{0}$-decay $\gamma$-ray spectrum in the Central Molecular Zone. Fig.~\ref{fig3}
shows this expectation as a grey shaded box. The measured data show a
significantly harder energy spectrum than this expectation, indicating
an excess of high energy hadrons in the GC, relative to the solar
neighbourhood. Interestingly, the spectrum of the diffuse emission is
very similar to that of the central source HESS J1745-290 (dashed line
in Fig.~\ref{fig3}).

From the lower panel of Fig.~\ref{fig2} it appears that the ratio of
$\gamma$-ray emission to molecular density varies with
galactic longitude, with a pronounced dip 
at $l\approx1.3^{\circ}$.  The implied non-uniform CR density
can be explained if CRs injected close to Sgr A have not yet had time
to diffuse out to $l\approx1.3^{\circ}$.
Assuming a diffusion coefficient of $D = 10^{30} \ \rm cm^{2} s^{-1}$,
the approximate value in the Galactic
Disk at TeV energies, an angular displacement of $1^{\circ}$ at the
distance of the GC corresponds to a diffusion time of $\sim$10$^{4}$
years, close to the age of the SNR Sgr~A~East~\cite{AEastAge}.
The energy required to fill the central 200 parsecs with CRs with the observed
density (extrapolating the measured H.E.S.S. spectrum down to 1~GeV)
is $10^{50}$ ergs, remarkably close to the CR energy input required from
a typical Galactic SNR in the paradigm of CR acceleration in these 
objects. A historical burst of activity in Sgr~A$^{\star}$ also 
provides a natural explanation for the observed CR excess.

\section{The Pulsar Wind Nebula in G0.9+0.1}
\label{secG09}

\begin{figure}[h]
\begin{minipage}{18pc}
\begin{center}
\includegraphics[width=17pc]{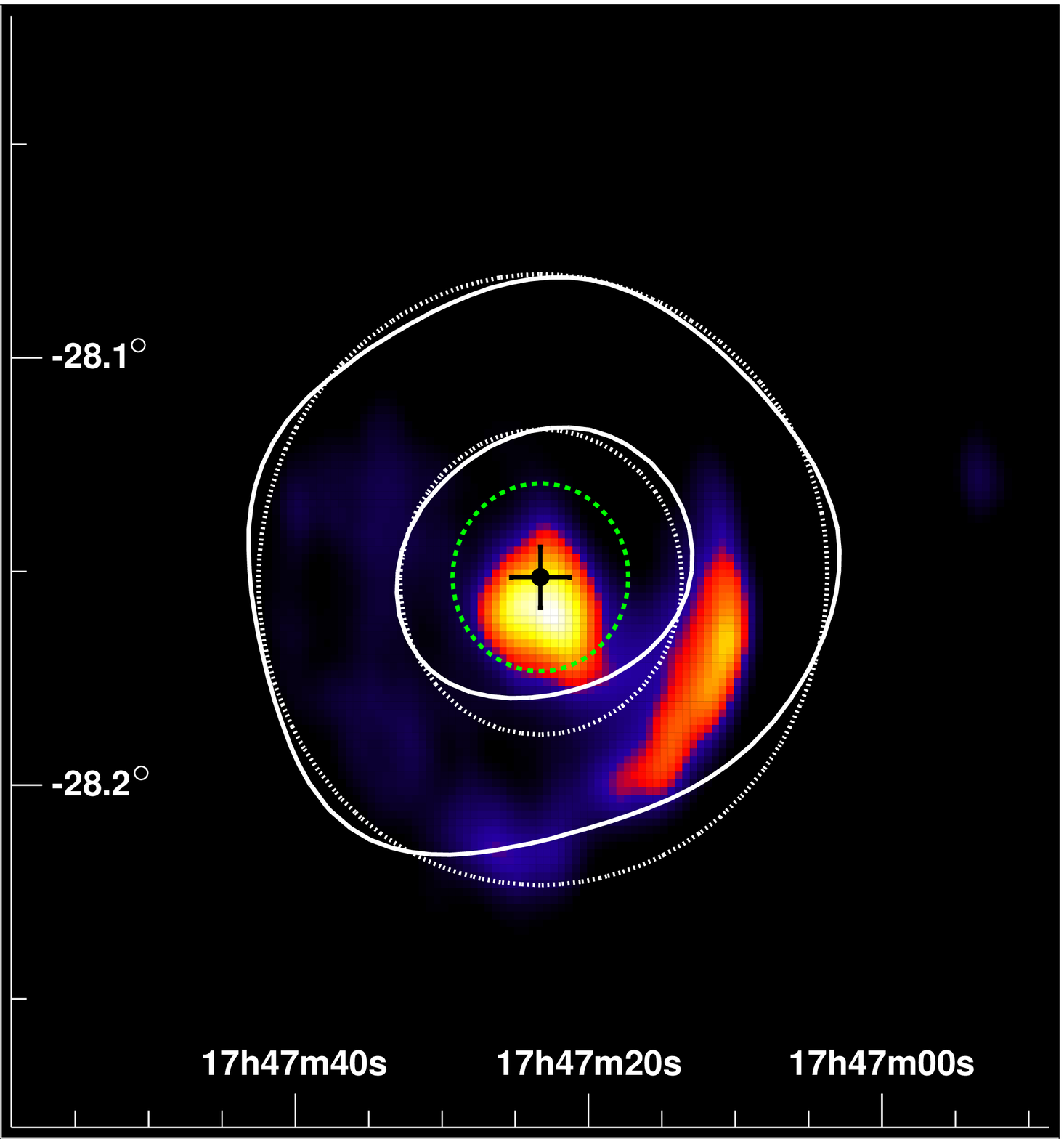}
\end{center}
\caption{\label{fig4} Comparison of the $\gamma$-ray source
  HESS\,J1747-281 with 90~cm radio emission from the composite SNR
  G0.9+0.1 (colour scale, \cite{LaRosa}).  The shape of the signal
  (solid contours) is compatible with the point-spread-function of the
  instrument (dotted contours).  The limit on the rms size of the VHE
  emission region is shown as a green dashed circle. The best fit
  position is marked with a cross (reproduced from \cite{HESS_g09}).
  }
\end{minipage}
\hspace{2pc}
\begin{minipage}{18pc}
\begin{center}
  \includegraphics[width=18pc]{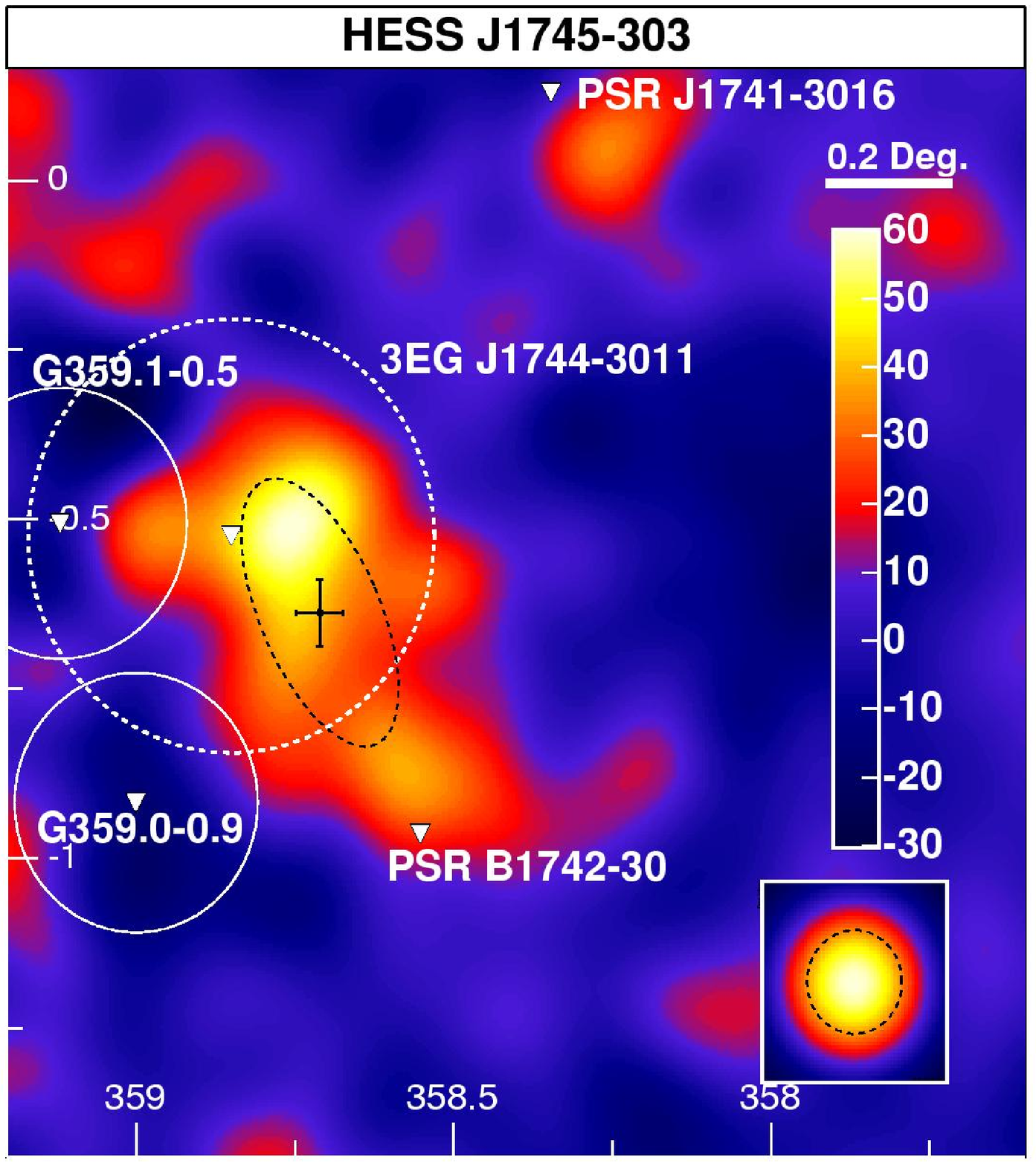}
\end{center}
\caption{\label{fig5} Smoothed and background subtracted $\gamma$-ray
  count map of the region around HESS\,J1745-303 in galactic coordinates. The best fit
  position and size of the H.E.S.S. source are marked with a black
  cross and a dashed ellipse. Nearby SNR and energetic pulsars are
  marked with triangles (reproduced from \cite{HESS_scan2}). The
  dashed white circle illustrates the 95\% confidence location error
  for the EGRET source 3EG\,1744-3011.}
\end{minipage}
\end{figure}

G\,0.9+0.1 is a composite radio SNR with a clear `core plus shell'
morphology (see for example~\cite{LaRosa}).  Observations with
Chandra~\cite{Chandra_g09} and XMM~\cite{XMM_g09} revealed hard
X-ray emission from the core and weak (likely thermal) emission from
the SNR shell.  The characteristic softening of the X-ray spectrum
with distance from the core indicates the likely plerionic nature of
this source, although no radio pulsar has been identified (likely due
to the large dispersion associated with the GC region). VHE
$\gamma$-ray emission coincident with G\,0.9+0.1 was discovered by the
H.E.S.S. collaboration in 2005~\cite{HESS_g09}.  Fig.~\ref{fig4} shows
the position of the H.E.S.S. source in comparison to radio
measurements.  The VHE emission is unresolved by H.E.S.S. and an upper
limit of $1.3'$ on the rms size of the emission region has been
derived. This limit allows us to exclude a $\gamma$-ray origin in the
shell of the SNR and strongly suggests a common plerionic origin of
the $\gamma$-ray and X-ray emission, arising respectively from inverse
Compton scattering and Synchrotron radiation of energetic electrons
within the nebula.

Pulsar wind nebulae (PWN) are prominent members of the VHE
$\gamma$-ray source catalogue. Recent $\gamma$-ray observations with
H.E.S.S. of known X-ray PWN such as Vela-X~\cite{HESS_velaX} and
MSH\,15-5{\it2}~\cite{HESS_1552} show close correlations between X-ray and VHE
$\gamma$-ray emission. Combined X-ray/$\gamma$-ray observations
provide information on the magnetic field in the nebula and on the
spatial distribution of electrons. For the distant PWN of G\,0.9+0.1
the emission is currently unresolvable in $\gamma$-rays but the
average magnetic field in the nebula can be estimated at $\sim
6\mu$G~\cite{HESS_g09}.

G\,0.9+0.1 is not seen in VHE $\gamma$-ray observations with MAGIC,
but the non-detection is not in conflict with the H.E.S.S. result,
given the lower sensitivity of the MAGIC dataset~\cite{MAGIC}.

\section{Unidentified Source HESS\,J1745-303}

The H.E.S.S. survey of the inner galaxy~\cite{HESS_scan2}, combined
with deep exposures on targets such as the GC and
RX\,J1713.7-3946~\cite{HESS_1713aa} has resulted in the detection of
many extended ($\sim10''$) $\gamma$-ray sources without clear
counterparts at other wavelengths.  One such source is
HESS\,J1745-303, located $\sim1^{\circ}$ from 
Sgr~A. Fig.~\ref{fig5} shows a count map of this source, illustrating
the positions of nearby supernova remnants and energetic pulsars.  The
most compelling potential counterpart is the (also unidentified) lower
energy $\gamma$-ray source 3EG\,1744-3011.  A fit of a elliptical
gaussian source profile yields a position of $l = 358.71 \pm
0.04^{\circ}, b = -0.64 \pm 0.05^{\circ}$ and an rms source size of
$5' \times 13'$. Deeper observations in all wavebands are clearly
desirable to help identify this MeV-TeV $\gamma$-ray source.

\section{Conclusions}

Present and future $\gamma$-ray observations will play a key role in
our understanding of the physical processes at work in the Galactic
Centre. Key outstanding issues, such as the spatial distribution and
strength of magnetic fields and the energy density of relativistic
particles, can be addressed with the help of such observations.
H.E.S.S. has provided the first sensitive view of this region in
very-high-energy $\gamma$-rays. In a few years time the combination of
the second phase of the H.E.S.S. project and the GLAST satellite will
provide unbroken sensitive coverage of the $10^{8}$--$10^{13}$ eV
$\gamma$-ray domain, with important consequences for Galactic Centre
research.


\end{document}